\documentclass[10pt]{article}
\usepackage{amsmath,amssymb,amsbsy,amsthm}

\let\epsilon\varepsilon
\let\phi\varphi

\newtheorem{theorem}{Theorem}

\newtheorem{definition}{Definition}

\begin{document}

\title{Using Information Theory to Study the Efficiency and Capacity  of   Computers   and Similar Devices}
\author{ Boris Ryabko${}^*$}
\date{}
\maketitle
\centerline{${}^*$ Siberian State University of Telecommunications and Informatics,}
\centerline{Institute of Computational Technologies of Siberian Branch of Russian Academy of Science,}
\centerline{Novosibirsk, Russia;   boris{@}ryabko.net }
\begin{abstract}
We address the problems of estimating  the  computer efficiency and the computer 
  capacity.   We define  the  computer efficiency and capacity   and   suggest
  a method for their estimation,  based on the analysis of    processor instructions and kinds of accessible  memory.
 It is shown  how the suggested  method can be  applied to  estimate the  computer capacity.  
In particular, this consideration gives a new look at the organization of the
 memory of a computer.  Obtained results can be of some interest for  practical applications. 
\end{abstract}
\vskip6mm \centerline{ {\bf Keywords:}  computer capacity,  computer efficiency,   Shannon theory,} 
\centerline{ cache memory, Information Theory}

\section{Introduction} 


We address the problem of what the  efficiency (or performance) and the capacity   of a computer are
 and how they can be estimated. 
More precisely, we consider  a computer with a certain set of instructions and several  kinds of  memory.  
What is the computer  capacity, if we know the execution time of each instruction and the speed of each kind of memory?  
What is the computer efficiency if the computer is used for solving problems of a certain kind (say, matrix multiplications)?
 On the one hand, the  questions about the computer efficiency and  capacity are quite natural, but,  on the other hand, to the best of our knowledge, the 
computer science does not give   answers to those questions.  

The first goal of this paper is to suggest a reasonable definition of the computer efficiency and  capacity    and   methods of their estimation. 
 We will mainly  consider computers, but our approach can be applied to all devices which  contain processors, memories and instructions. 
 (Among those devices we mention mobile telephones and  routers.)
  Second, we describe a method for estimation of the computer capacity  and apply it to several examples which are of some theoretical and practical  interest.
  In particular, this consideration gives a new look at the organization of  a computer memory. 

The suggested 
 approach is based on the concept of  Shannon entropy, the capacity of a discrete noiseless channel and some other  ideas of C.~Shannon  \cite{S}
that underly Information Theory.

\section{The computer efficiency and   capacity}

\subsection{The basic concepts and definitions}
Let us first  briefly describe the main point of the suggested 
approach and definitions. For a start, we will consider the 
simplified variant of a computer,  which consists of a set 
of instructions $I$  and an accessible memory $M$. 

We suppose that at the initial moment there is a program and 
data which   can be considered as  binary words $P$ and $D$,
  located in the memory of a computer $M$.
In what follows we will call the pair $P$ and 
$D$  a computer task.
 A computer task $<P,D>$   determines  a certain sequence of instructions 
$X(P,D)$ $=x_1 x_2 x_3 ... ,\, \,  x_i \in I$. (It is supposed that an
 instruction may contain an address of 
a memory location in $M$,  the  index of a register, etc.)
 For example, if the program $P$ contains a loop  which will be 
executed   ten times, then the sequence $X$ will contain the 
body of  this loop repeated ten  times. 
 We say that two computer tasks 
$<P_1,D_1>$ and $<P_2,D_2>$ are different, if the sequences $X(P_1,D_1)$ and  $X(P_2,D_2)$ 
are different. 

Let us   denote the execution time of an instruction $x$ by $\tau(x)$.
Then  the execution time $\tau(X)$ of a sequence of instructions 
$X$ $=x_1 x_2 x_3 ... x_t $ is given by
$$\tau(X) = \sum_{i=1}^t \tau(x_i) .
$$ 
The key observation is as follows: the number of different computer tasks, whose execution time equals $T$, is upper bounded by  
the size of the  set of all sequences of instructions, whose  execution time equals $T$, i.e.
\begin{equation}\label{nu}
 \nu(T)  \leq   \,  N(T),
\end{equation}
where $ \nu(T) $ is the number of different problems, whose execution 
 time equals $T$, and 
\begin{equation}\label{N}
N(T) =|  \{  X:  \tau(X) =  T \} | .
\end{equation} 

Hence, 
\begin{equation}\label{log}
\log  \nu(T) \leq  \log N(t). 
\end{equation}
(Here and below $\log x \equiv \log_2 x$ and  $|Y|$ is the number of elements of $Y$ 
if $Y$ is a set, and the length of $Y$ if $Y$ is a word.)
In other words, the total number of computer 
tasks
executed in  time $T$ is upper bounded by 
(\ref{N}).

Basing on this consideration we give the following definition.
\begin{definition}
 Let there be a computer  with a set 
of instructions $I$   and let $\tau(x)$ be the execution time of an instruction $x  \in I.$ 
The computer capacity $C(I)$ is defined as follows:
\begin{equation}\label{cap}
C(I) \, = \, \lim_{T \rightarrow\infty}  \frac{  \log N(T)} { T} \, ,
\end{equation}
 where $N(T)$ is defined in (\ref{N}).
\end{definition}

(That this limit always exists can be proven  based on the lemma by M.~Fekete  \cite[lemma M.~Fekete]{kr}.)

The next question to be investigated is the definition of the computer efficiency 
(or performance), when a computer is used for solving  problems of a certain 
kind.  For example, one computer can be a Web server, another can be used for solving differential equations, etc. Certainly, the computer efficiency 
 depends on the problems the computer has to solve. In order to model this situation 
  we suggest the following approach: there is an information source which generates a sequence of computer tasks
in such a way, that the computer begins to solve each next task as soon as  the previous task is finished.
 We will not deal with a probability distribution on the sequences of the computer tasks, but consider  sequences of  computer instructions, 
 determined by sequences of the computer tasks, as a stochastic processes.  
In what follows we will consider the model when this stochastic process is stationary and ergodic, and we will define the computer efficiency for this case. 

The definition of  efficiency will be based on results and ides of information theory, which we introduce in what follows.
Let there be a stationary and ergodic process $ z =
z_1, z_2, ...
$ generating letters from a finite alphabet $A$ (the definition of stationary 
argodic process can be found, for ex., in \cite{co}).
The $n-$order Shannon entropy and the limit Shannon entropy are defined as follows:
\begin{equation}\label{ent}
 h_n(z) = -  \frac{1}{n+1}\sum_{u \in A^{n+1} } P_z(u) \log P_z(u) ,  \, \, 
\, \, h_\infty(z) = \lim_{n \rightarrow \infty} h_n(z)
\end{equation}
where 
$n \ge 0$ , $P_z(u)$ is the probability that $z_1 z_2 ... z_{|u|} $ $ = u$ (this limit always exists, see  \cite{co,S}).
We will consider so-called  i.i.d. sources. By definition, they
generate independent and identically distributed random variables from some set
$A$.
Now we can define the computer efficiency.  
\begin{definition}  
 Let there be a computer  with a set 
of instructions $I$   and let $\tau(x)$ be the execution time of an instruction $x  \in I.$ 
 Let this computer be used for solving such a randomly generated sequence of computer tasks,
that the corresponding  sequence of the instructions $z = z_1 z_2 ... $, $ z_i \in I$,
is a stationary ergodic stochastic process. Then  the efficiency is defined as follows:
\begin{equation}\label{eff}
c(I,z) = h_\infty(z) / \sum_{x \in I} P_z(x) \tau(x) ,
\end{equation}
where $P_z(x) $ is the probability that $z_1=x, x \in I$.
\end{definition}
 Informally, the Shannon entropy is a quantity of information (per letter),
 which can be transmitted and the denominator in (\ref{eff}) is the average execution time of an instruction.  

More formally, if we take a
 large
integer $T$ and consider  all $T-$letter sequences
$z_1 ... z_T$, then, for large $T$, the number of ``typical''  sequences  will be approximately $2^{T  h_\infty(z)}, $ whereas  
the total execution time of the sequence will be approximately $T \sum_{x \in I} P_z(x) \tau(x).$ (By definition of a typical sequence, the
  frequency  of any word $u $ in it is close to the probability $P_z(u)$. The total probability of the set of all typical sequences is close to 1.)  So, 
the ratio of  $ \log (2^{T  h_\infty(z)})  $ and the average execution time will be asymptotically equal to (\ref{eff}), if 
$T \rightarrow  \infty.$  A rigorous proof can be obtained  basing on  
methods of information theory; see \cite{co}. We do not give it, because definitions do not need to be proven,
but mention that there are many  results about 
channels which transmit letters of unequal duration \cite{cs,kra,me}. 

\subsection{Methods for estimating the computer capacity }

        Now we consider the question of estimating the  computer capacity and efficiency defined above. 
The efficiency, in principle, can be estimated basing on statistical data, which 
 can be obtained by observing a computer which solves  tasks of a certain kind.

The computer capacity $C(I)$ can be estimated in different situations by different methods.
In particular, a stream of instructions generated by different computer
 tasks can be described as a sequence of words created by a formal language, or
the dependence between sequentially  executed instructions can be modeled by Markov chains, etc. 
Seemingly the most general approach is to define the set of admissible sequences of instructions 
as a certain subset of all possible sequences. 
 More precisely,  the set of admissible sequences   $G$ is defined 
as a subset $G \subset A^\infty, $ where $A^\infty$ is the set of one-side infinite words
over the alphabet $A$: $A^\infty = \{x: x = x_1 x_2 ... \},$ 
$x_i \in A, i=1,2, ...  .$
In this case the the capacity of $G$ 
is deeply connected with the topological entropy and Hausdorff dimension;
for definitions and examples see \cite{an,do,fo,ry} and references therein. 
We do not consider this approach in details, because it seems to 
be difficult to use it for solving applied problems 
which require a finite description 
of the channels.

The simplest  estimate of  computer capacity  
can be obtained if we suppose that all sequences of the instructions 
are admissible.  In other words, we consider the set of instructions
$I$ as an alphabet and suppose that 
all sequences of letters (instructions) can be executed. 
In this case the method of calculation of the lossless channel  capacity, given by C.Shannon in \cite{S},
 can be used. It is important to note that this method can be used  
for upper-bounding the computer capacity for all other models, because for any computer the set of
admissible sequences of instructions is a subset of all words over the "alphabet" $I$.

Let, as before, there be a computer with a set 
of instructions $I$   whose execution time is $\tau(x),  x  \in I,$ 
  and  all  sequences of instructions are allowed. In 
other words, if we consider the set $I$ as an alphabet, then all possible 
words over this alphabet can be considered as 
admissible sequences of  instructions for the computer. 
The question we consider now is how one can calculate (or estimate) the  
capacity (\ref{cap}) for this case.  The solution is suggested by 
C.~Shannon \cite{S} who  showed that the capacity $C(I)$ is equal to the logarithm of the largest real 
solution $X_0$ of the following equation:
 \begin{equation}\label{X}
X^{-\tau(x_1)} + X^{-\tau(x_2)} + ... +X^{-\tau(x_s)}  = 1, 
\end{equation} 
where $I = \{ x_1, ... , x_s \}$.  In other words, $C(I) = \log X_0 . $

It is easy to see that the efficiency  (\ref{eff})  is maximal, if 
the sequence  of instructions $ x_1 x_2 ... $, $x_i \in I$ 
is generated by an i.i.d. source with probabilities  
$p^*(x)   = X_0^{- \tau(x)}, $    where $X_0$  is the largest real solution to 
the equation (\ref{X}), $x \in I$. Indeed, 
having taken into account that $h_\infty(z) = h_0(z)$ for i.i.d. source \cite{co} and  the definition of entropy
(\ref{ent}),  the direct calculation of $c(I,p^*) $ 
in (\ref{eff}) shows that  $c(I,p^*) = $ $ \log X_0$ and, hence, 
$c(I,p^*)  = C(I).$

It will be convenient to combine all the results about computer capacity 
and efficiency in the following statement: 

\begin{theorem}\label{TR}
Let  there be a computer with a set 
of instructions $I$ and let $\tau(x)$ be the execution time of  $ x  \in I$.
Suppose  that
   all  sequences of instructions are admissible computer programs.  
  Then 
 the following equalities are valid:
\begin{itemize}
 \item[i)]   The alphabet capacity $C(I)$   (\ref{cap}) equals $\log X_0,$ 
where $X_0$  is the largest real solution to 
the equation (\ref{X}). 
 \item[ii)] The efficiency  (\ref{eff})  is maximal if the sequences of instructions are generated by an i.i.d. source 
with probabilities  
$p^*(x) = X_0^{- \tau(x)},$ $x \in I.$
\end{itemize}
\end{theorem}

\section{MIX and MMIX}
As an example we briefly  consider the MIX and MMIX computers suggested by D.Knuth \cite{kn1,kn2}. The point 
is that those computers
are described in details and MMIX can be considered as  a model of a modern
computer, whereas  MIX can be considered as a model of computers produced in the 1970th. 
The purpose of this consideration is to investigate the given definitions and to look at how 
various characteristics of a computer influence its capacity, therefore we give some details of the 
description of MIX and MMIX.

We consider a binary version of MIX \cite{kn1}, whose instructions are represented by $31-$bit words. 
Each machine instruction occupies one word in the memory, and consists of 4 parts:
 the address (12 bits and the sign of the word) in memory to read or write; an index specification 
(1 byte, describing which  register to use) to add to the address; a modification (1 byte) that specifies 
which parts of the register or memory location will be read or altered; and the operation code (1 byte). 
 So, almost all $31-$bit words can be considered as possible instructions and the upper bound of the number of the set of instructions
$I$  (and   letters of the 
"computer alphabet")  is $2^{31}.$
Each MIX instruction  has an associated execution time, given in arbitrary units.  For example, the instruction
 $JMP$ (jump) has the execution time 1 unit, the execution times of   $MUL$ and $DIV$  
(multiplication and division) are 10 units and 12 units, correspondingly. 
There are special instructions whose execution time 
 is not constant. For example, the instruction  $MOVE$ is intended 
to copy  information from several cells of memory and the execution time equals $1 + 2 \,  F$,
where  $F$  is the number of cells.  

 From the description of MIX instructions and Theorem 1 we obtain the following equation for calculating
 the upper bound of the  capacity of MIX:
\begin{equation}\label{ex1}
  \frac{2^{28} } {X} +  \frac{2^{26} } {X^2 } +  \frac{2^{26} } {X^{10}}  +  \frac{2^{25} } {X^{12}}   + \sum_{F=0}^{2^{25}}  \frac{2^{25} } {X^{1+ 2 F}}         \, = \, 1  .
\end{equation} 
Here the first summand corresponds to operations with execution time 1, etc.  It is easy to see that
the last sum can be estimated as follows:  $ \, \sum_{F=0}^{2^{25}}  \frac{2^{25} } {X^{1+ 2 F}}  < \frac{2^{25}}{X}   \frac{X^2}{X^2 - 1} .$ 
Having taken into account this inequality and (\ref{ex1}), we can obtain by  direct calculation that 
the    MIX capacity is approximately 28 bits per  time unit. 

The MMIX computer has 256 general-purpose registers, 32 special-purpose ones and $2^{64}$ bytes of 
virtual memory \cite{kn2}.  The MMIX instructions are presented as 32-bit words and in this case the "computer alphabet" 
 consists of almost $2^{32}$  words (almost, because some combinations of bits do not make sense).  
In \cite{kn2} the execution (or running) time is assigned to each instruction in such a way that each instruction 
takes an integer number of $\upsilon$, where $\upsilon$ is a unit that represents the clock 
cycle time.  Besides, it is assumed that the running time depends on the number of memory references 
(\emph{mems})
that a program uses \cite{kn2}.  For example, it is assumed that the execution time of  
each of the $LOAD$ instructions is $\upsilon + \mu$, where $\mu$ is an average time of memory 
reference \cite{kn2}.  If we consider $\upsilon$ as the time unit and define $\hat{\mu} = \mu/ \upsilon$,
 we obtain from  the  description of MMIX  \cite{kn2} and Theorem 1  the following equation for finding an upper bound on the MMIX capacity:
\begin{equation}\label{ex2}
   2^{24} \,  ( \,  \frac{139} {X}  +  \frac{32} {X^{2}}  +   \frac{5} {X^3} +  \frac{17} {X^4} +  \frac{3} {X^5} +
 \frac{4} {X^{10}} +  \frac{2} {X^{40}}  +  \frac{4} {X^{60}} + $$ $$
  \frac{46} {X^{1+\hat{\mu}}}  +
 \frac{2} {X^{1+ 20 \hat{\mu}}}  +
 \frac{46} {X^{2+ 2 \hat{ \mu}}}  \,
  )
  \, = \, 1  . 
 \end{equation}  
The value $\hat{\mu}$ depends on the realization of MMIX and is larger than 1 for modern computers \cite{kn2}.
So, as in the previous example, the first term has the most influence and   
 MMIX capacity is approximately 31.5 bits per  time unit. 

These examples show that the capacity of both computers is mainly determined by the subsets of
 instructions whose execution time is minimal.   

Theorem 1 gives a possibility to estimate 
frequencies of the instructions, if the computer performance efficiency is maximal (and equals its capacity). 
First, the frequencies of instructions with equal running time have to be equal. 
In turn, it means that all memory cells should be used equally often. Second, the frequency of instructions  exponentially decreases 
as their running time increases.  
It is interesting that  in   the modern computer MMIX the share of fast commands 
is larger than in  the  old computer MIX and, hence, the efficiency of MMIX is larger. 
It is reached due to the usage of registers  instead of  the (slow) memory.

\section{Possible applications}

It is natural to use estimations of the computer capacity  at the  design stage. 
We consider examples of such estimations that are intended to illustrate some possibilities of the suggested approach.

First we consider a computer, whose design is close to the MMIX computer.  Suppose a designer has decided  to   use the MMIX
set of registers. Suppose further, that he/she  has a possibility to use two different kinds of memory,
 such that the time of one reference to the memory and the cost of one cell 
are $\tau_1,$ $c_1$ and  $\tau_2,$ $c_2,$ correspondingly. It is natural to suppose that the total price  of the memory is required
 not  to exceed a certain bound $C$. As in the example with MMIX we define 
$\hat{\mu}_1 = \tau_1 / \upsilon$,  $\hat{\mu}_2 = \tau_2 / \upsilon$, where, as before, $\upsilon$ is a unit that represents the clock 
cycle time. 

As in the case of MMIX, we suppose that there are   instructions 
 for writing and reading information from a  register 
 to  a cell. The set of these instructions 
coincides with the corresponding set of the MMIX computer. 
   If we denote the number of  the memory cells by $S$, then
   the number of the instructions which can be used for reading and writing,   is proportional to  $ S$. 
Having taken into account that MMIX has $2^8$ registers and the equation (\ref{ex2}) , we can see that 
  the designer should consider
two  following  equations 
\begin{equation}\label{ex3}   (2^{24} \,  ( \,  \frac{139} {X}  +  \frac{32} {X^{2}}  +   \frac{5} {X^3} +  \frac{17} {X^4} +  \frac{3} {X^5} +
 \frac{4} {X^{10}} +  \frac{2} {X^{40}}  +  \frac{4} {X^{60}}  \,) )+ $$ $$ 2^8 \, S_i  \, (
  \frac{46} {X^{1+\hat{\mu}_i}}  +
 \frac{2} {X^{1+ 20 \hat{\mu}_i}}  +
 \frac{46} {X^{2+ 2 \hat{ \mu_i}}}  \,
  ) \, ) 
  \, = \, 1  . 
\end{equation}
 for $i=1,2$, where $S_i = C/c_i$, i.e. $S_i$ is the number of cells of the $i-$th kind of  memory,
 $i=1,2$.  The designer can 
 calculate the maximal roots for each  equation ($i =1, 2$)  and then he/she can choose that kind of memory for which 
the solution is larger.
It will mean that the computer capacity will be larger for the chosen kind of memory.  
For  example,  suppose that the total price should not exceed $1$ ($C=1$), 
the prices of one cell of memory are $c_1 = 2^{-30}$ and $c_2 = 2^{-34}$, whereas $\hat{\mu}_1= 1.2$, $ \hat{\mu}_2 = 1.4$.
  The direct calculation of the equation (\ref{ex3}) for $S_1 = 2^{30}$  and  $S_2 = 2^{34}$
shows that the former is  preferable, because the computer capacity is larger for the first kind of memory. 

Obviously, this model can be generalized for different set of instructions and different kinds of memory. In such a case 
the considered problem can be described as follows.  We suppose that there are   instructions 
$\mu^w_i(n)$ for writing information from a special register 
 to 
$n$-th cell of $i$-th kind of memory ($n=0,... , n_i - 1,$ $1=1, ... , k,$
and similar instructions $\mu^r_i(n)$ for reading. Moreover, it is supposed that all other instructions cannot directly 
read
or write  to the memory of those kinds, i.e. they can write to and 
read from the registers
only. (It is worth noting that this model is quite close to
some real computers.) 
Denote the execution time of the instructions $\mu^w_i(n)$ 
and $\mu^r_i(n)$ by $\dot{\tau}_i, \, 1=1, ... , k.$

In order to get an upper bound of the computer capacity for the described model
we, as before, consider the set of instructions as an alphabet and estimate  its 
capacity
applying Theorem 1. From (\ref{X}) we obtain that the capacity 
is 
$\log X_0$, where $X_0$ is the largest real solution of the 
following equation:\begin{equation}\label{mem}
\sum_{x \in I^* }X^{-\tau(x)} +
R \, ( \frac{2 n_1}{X^{\dot{\tau}_1}}
+\frac{2 n_2}{X^{\dot{\tau}_2}} + ... +
\frac{2 n_k}{X^{\dot{\tau}_k}} )\, = \, 1 \, ,
\end{equation} 
 where $I^*$ contains all instructions except 
$\mu^r_i(n)$ and $\mu^w_i(n) , \, 1=1, ... , k, $ 
$R$ is a number of registers. (The summand
$\frac{2 n_i}{X^{\dot{\tau}_i}}$ corresponds to the instructions
$\mu^w_i(n)$ and $\mu^r_i(n)$.) 

Let us suppose that a price of one cell of $i$th kind of memory is $c_i$   
whereas the total cost of memory is limited by $C$. Then, from the previous equation we obtain the following optimization problem:
$$
  \log X_0  \longrightarrow  maximum,
$$
where $X_0$ is the maximal real solution of the equation (\ref{mem}) and 
$$ c_1 n_1 + c_2 n_2 + ... + c_k  n_k  \, \le  \, C ; \, n_i  \ge 0, \, i = 1, ... , k .$$
 The solution of this problem can be found using standard methods and used by computer  designers.

The suggested approach can be applied to optimization of 
different parameters of computers including the structure of the set of instructions, etc. 

\section{Conclusion}

We have suggested a definition of the computer capacity and its efficiency as well as  a method for their estimation.
It can be suggested that this approach may  be useful on the design stage  when developing computers and similar devices. 

It  would be interesting to analyze the ``evolution'' of computers from the point of view of
 their capacity.  The  preliminary analysis shows that   the development of the RISC  processors,
the increase in quantity of the registers and some other innovations, lead  to the  increase of the capacity of  computers.
Moreover, such methods as using  cache memory can be interpreted as an attempt to increase the efficiency of a computer.

It is worth noting that the suggested   approach in general can be extended 
to multi-core processors and special kinds of cache memory.

\end{document}